\documentclass{elsart1p}
\usepackage{graphicx}
\usepackage{amssymb}
\newcommand{\dn}{\frac{dN^\phi}{dy}} 
\newcommand{\pt}{\langle p_T^\phi\rangle}
\newcommand{\vih}{\left (\frac{v_2}{\epsilon} \right )^{ih}}
\newcommand{\etas}{\left ( \frac{1}{\tau_i T_i} \frac{\eta}{s} \right )}

\begin{document}
\begin{frontmatter}
%
%
%
%
%

\title{Causal dissipative hydrodynamics for heavy ion collisions }
%
%

\author{A.K. Chaudhuri}

\address{Variable Energy Cyclotron Centre\\1-AF, Bidhan Nagar, Kolkata-700 064 }
\ead{akc@veccal.ernet.in}
\begin{abstract}
We briefly discuss the recent developments in causal dissipative hydrodynamic for relativistic heavy ion collisions. 
Phenomenological estimate of QGP viscosity over entropy ratio from several experimental data, e.g. STAR's $\phi$ meson data, centrality dependence of elliptic flow, universal scaling elliptic flow etc. are discussed. 
QGP viscosity, extracted from hydrodynamical model analysis can have very large systematic uncertainty due to uncertain initial conditions.  
\end{abstract}
\end{frontmatter}

\section{Introduction} \label{sec1}
In recent years there is much interest in QGP viscosity. In any fluid motion, some amount of irreversibility is present that leads to energy dissipation. It is then important to have estimates of dissipative coefficients like viscosity of QGP.  Without this estimate, it is difficult to claim that the matter produced in RHIC collisions can be understood, at least in part, with thermodynamic concepts.
Theoretically,    it is possible to estimate dissipative coefficients of   QGP from first principle, however, the problem is complex and yet unsolved. One hopes to extract viscosity from experimental data.  
It require numerical implementation of dissipative hydrodynamics.  
In recent years significant progress has been made in  numerical implementation 
of dissipative hydrodynamics 
\cite{Chaudhuri:2006jd,Chaudhuri:2008sj,Chaudhuri:2005ea,Chaudhuri:2008ed,Song:2008hj,Song:2009rh,Romatschke:2007mq,Luzum:2008cw,Dusling:2007gi}.
In the following, we briefly describe causal dissipative hydrodynamics. We also discuss about some estimates of shear viscosity to entropy ratio of QGP from experimental data.

\section{Causal dissipative hydrodynamics} \label{sec2}
Relativistic generalisation of Navier-Stokes equation, called 1st order theories \cite{Eckart,LL63} for dissipative hydrodynamics, suffer from the problem of acausality and instability \cite{Hiscock:1985zz}. The problem is eliminated 
in Israel-Stewart's \cite{IS79} 2nd order theory (for detailed exposition see
\cite{Muronga:2003ta,Heinz:2005bw}). Briefly,   dissipative fluxes  are treated as extended thermodynamic variables and relaxation equations for the dissipative fluxes  are obtained from the entropy law; $\partial_\mu S^\mu \geq 0$. Space-time evolution of the fluid is then obtained by solving energy-momentum conservation equations and relaxation equations for dissipative fluxes,
   
\begin{eqnarray}  
\partial_\mu T^{\mu\nu} & = & 0,  \label{eq1a} \\
D\pi^{\mu\nu} & = & -\frac{1}{\tau_\pi} (\pi^{\mu\nu}-2\eta \nabla^{<\mu} u^{\nu>})  
-[u^\mu\pi^{\nu\lambda}+u^\nu\pi^{\mu\lambda}]Du_\lambda \label{eq1b} \\
D\Pi&=&-\frac{1}{\tau_\Pi}(\Pi+\zeta \partial_\mu u^\mu) \label{eq1c} 
\end{eqnarray}

Eq.\ref{eq1a} is the conservation equation for the energy-momentum tensor, $T^{\mu\nu}=(\varepsilon+p)u^\mu u^\nu - (p+\Pi)g^{\mu\nu}+\pi^{\mu\nu}$, 
$\varepsilon$, $p$ and $u$ being the energy density, hydrostatic pressure and fluid 4-velocity respectively.    
In Eq.\ref{eq1b}, $D=u^\mu \partial_\mu$ is the convective time derivative, $\nabla^{<\mu} u^{\nu>}= \frac{1}{2}(\nabla^\mu u^\nu + \nabla^\nu u^\mu)-\frac{1}{3}  
(\partial . u) (g^{\mu\nu}-u^\mu u^\nu)$ is a symmetric traceless tensor. 
 Eq.\ref{eq1b} and \ref{eq1c} are the relaxation equations for the shear stress tensor $\pi^{\mu\nu}$ and bulk  pressure $\Pi$, with $\tau_\pi$ and $\tau_\Pi$ as the relaxation times.
$\eta$ and $\zeta$ are shear and bulk viscosity coefficients. We note that relaxation equations for dissipative fluxes are non-trivial  and may contain additional terms \cite{Betz:2009zz,Betz:2008me}. Eqs.\ref{eq1a}-\ref{eq1c} are closed only with an equation of state (EOS) $p=p(\varepsilon)$. Any phase transition in the system can be incorporated through the EOS. 
Boost-invariant version of Eqs.\ref{eq1a}- \ref{eq1c} has been solved by many authors \cite{Chaudhuri:2006jd,Chaudhuri:2008sj,Chaudhuri:2005ea,Chaudhuri:2008ed,Song:2008hj,Song:2009rh,Romatschke:2007mq,Luzum:2008cw,Dusling:2007gi}\footnote{In \cite{Dusling:2007gi} relaxation equations are obtained in {\"Ottinger} and Gremala formulation  \cite{Grmela:1997zz}.}. Solution require initial conditions e.g. the initial or the thermalisation time, initial energy density, fluid velocity distribution. In viscous hydrodynamics additionally, shear stress tensors and bulk pressure has to be initialised. A priori, they are unknown and only way to fix them is to confront the theory with experimental data. Fitting large number of parameters with experimental data is a complex process. 
To reduce free parameters, some of the parameters are fixed, e.g.
initial transverse energy density/fluid velocity profile is fixed from Glauber model or color glass condensate model, initial shear stress tensors are fixed either to zero or boost-invariant values etc. see \cite{Chaudhuri:2006jd,Chaudhuri:2008sj,Chaudhuri:2005ea,Chaudhuri:2008ed,Song:2008hj,Song:2009rh,Romatschke:2007mq,Luzum:2008cw,Dusling:2007gi}. All the parameters are not independent either. For example, initial time and central energy density are inversely related. Viscosity flattens $p_T$ spectra, an effect that can be mimicked by initial fluid velocity. Hydrodynamic models also require a freeze-out prescription. In ideal hydrodynamics, one generally use sudden freeze-out, where on some assumed freeze-out surface hydrodynamically evolving fluid suddenly transforms in to free-streaming particles. Most of the viscous hydrodynamics also uses the same algorithm. Recently, Dusling and Teaney \cite{Dusling:2007gi} implemented dynamical freeze-out through the freeze-out condition, 
$\tau_{rel} \partial u \approx 0.5$. It satisfies the viscous hydrodynamics requirement that microscopic relaxation time is much smaller than the macroscopic
inverse expansion rate.

\section{Hydrodynamical model's estimates of $\eta/s$ from experimental data}\label{sec3}
As mentioned earlier, one of the aims of causal hydrodynamics is to extract QGP viscosity from experimental data.
In \cite{Luzum:2008cw}, charged particles multiplicity, radial flow and elliptic flow in $\sqrt{s_{NN}}$=200 GeV Au+Au collisions were analysed to obtain an estimate of $\eta/s$. They also estimated systematic uncertainty both in theory and experiments, $\eta/s=0.1\pm 0.1 (theory) \pm 0.08 (experiment)$.
Considering several deficiencies of the model, it was concluded that
QGP viscosity, $\eta/s \geq 0.5$ can be excluded. A similar conclusion was drawn in  \cite{Song:2008hj} from the analysis of elliptic flow data. In the following, we discuss in brief some recent estimates of $\eta/s$ from $\phi$ meson data, centrality dependence and scaling violation of elliptic flow.
 
\begin{figure}[ht]
\begin{minipage}{13pc} 
\resizebox{0.7\textwidth}{!}{\includegraphics{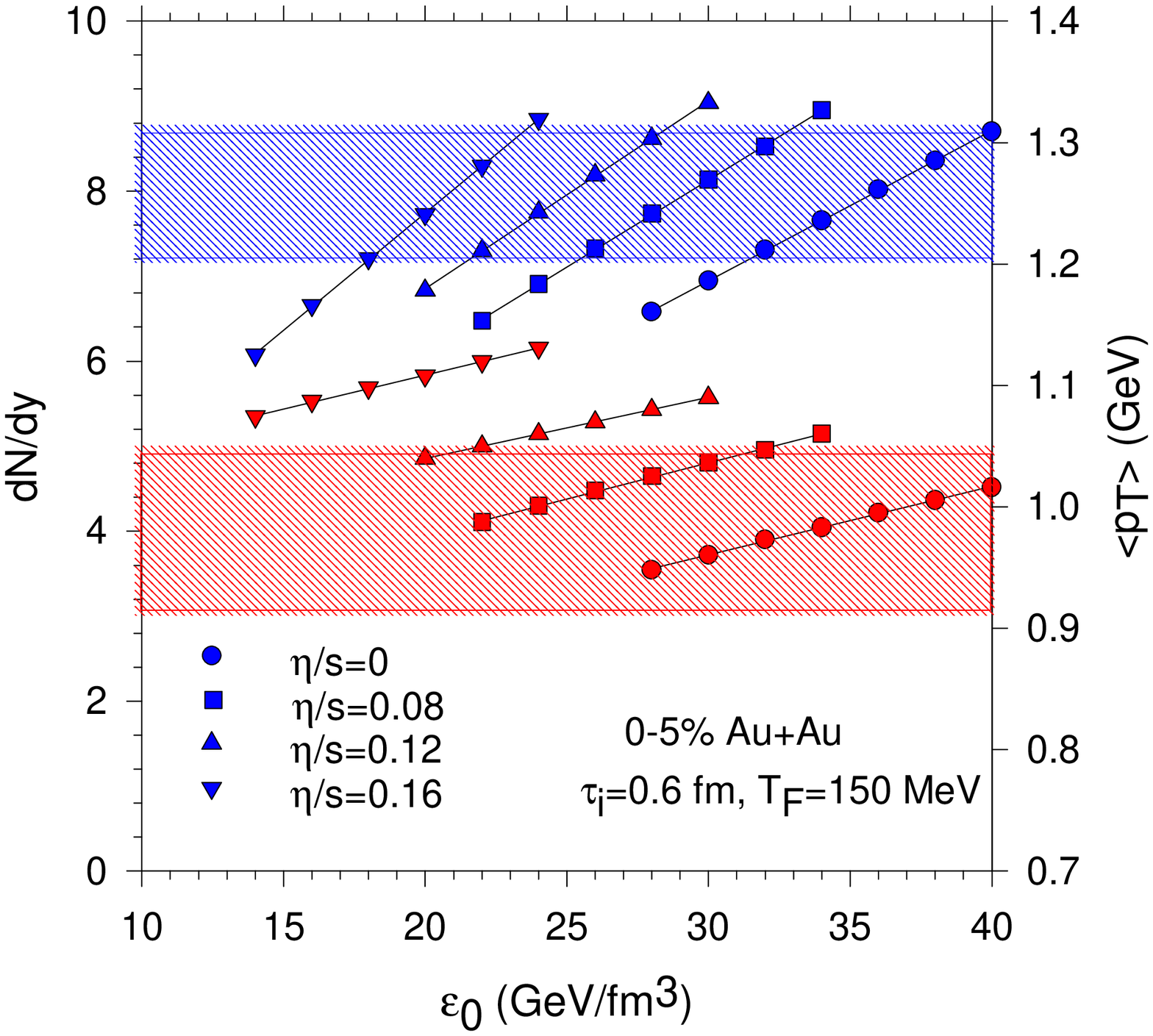}} 
 \end{minipage} \hspace{-3pc}
\begin{minipage}{15pc} 
\resizebox{0.50\textwidth}{!}{\includegraphics{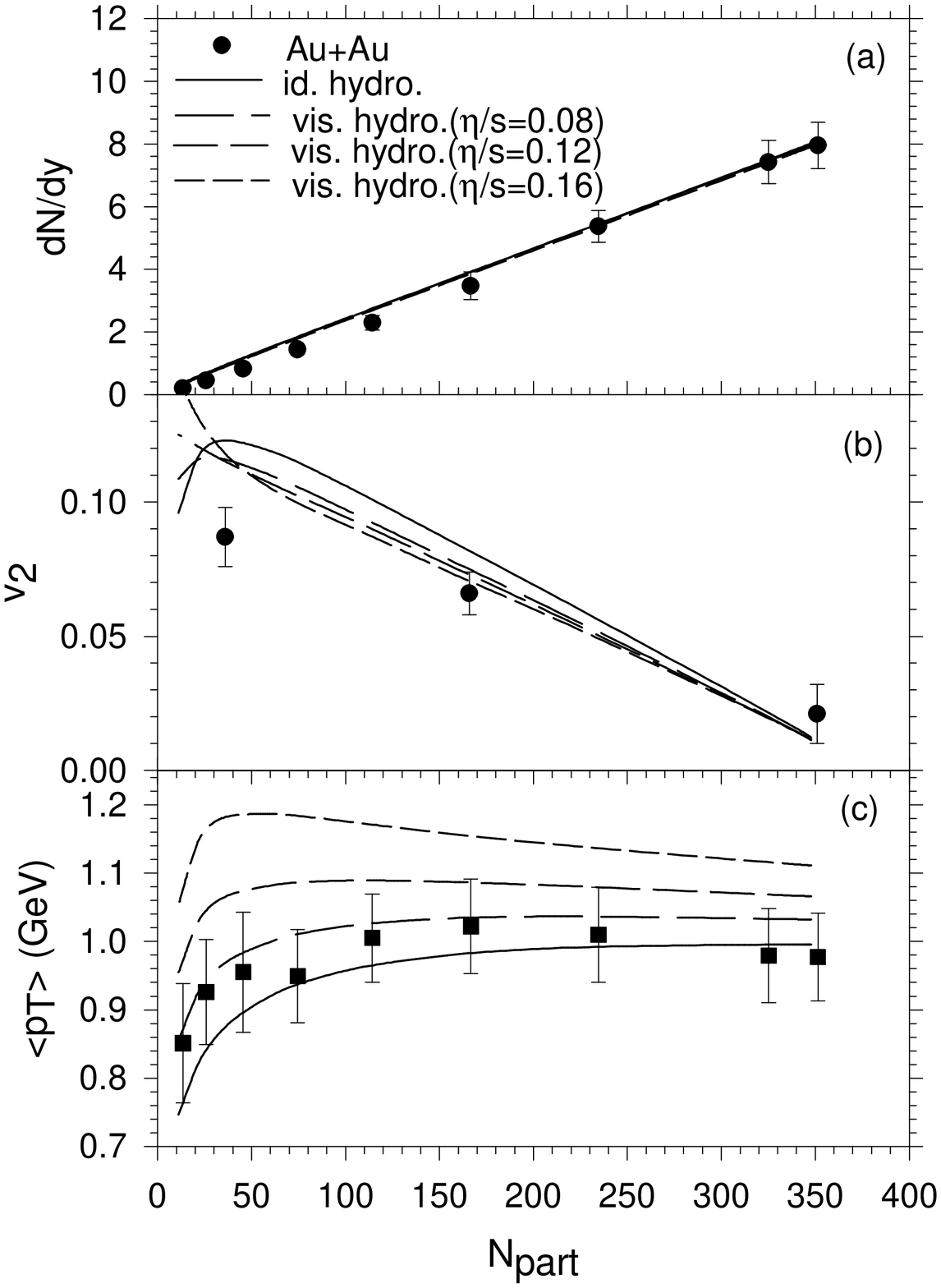}} 
 \end{minipage}\hspace{-6pc}
 \begin{minipage}{12pc}
\caption{ \label{F1} (color online) (left panel) Variation of $\dn$ and $\pt$ with central energy density in 0-5\% Au+Au collisions. (right panel) centrality dependence of $\phi$ meson multiplicity, integrated $v_2$ and  mean $p_T$
are compared with hydrodynamical simulations with central energy density as in table.\ref{table1}. Initial time $\tau_i$=0.6 fm, freeze-out temperature $T_F$=150 MeV. 
}
\end{minipage}
\end{figure}

 \begin{table}[ht] 
\begin{minipage}{20pc}
  \begin{tabular}{|c|c|c|c|c|}\hline
$\eta/s$         & 0    & 0.08 & 0.12 & 0.16 \\  \hline
$\varepsilon_i (GeV/fm^3$) & $35.5\pm 5.0$ & $29.1\pm 3.6$ & $25.6\pm 4.0$ &  $20.8\pm 2.7$ \\ \hline  
$T_i$ (MeV)  & $377.0\pm 13.7$ & $359.1\pm 11.5$ & $348.0\pm 14.3$ & 
$330.5\pm 11.3$     \\ \hline
\end{tabular}
\end{minipage} 
\begin{minipage}{10pc}
 \caption{\label{table1} Initial energy central energy density ($\varepsilon_i$) and temperature ($T_i$) from fit to $\phi$ multiplicity in 0-5\% Au+Au collisions.
} 
\end{minipage}
\end{table}  

\subsection{STAR's $\phi$ meson data and QGP viscosity}\label{sec4}

$\phi$ mesons have several unique features, e.g.
hidden strangeness, both hadronic and leptonic decay, not affected by resonance decays,   mass and width are not modified in a medium    etc. which  make them an ideal probe to investigate medium properties in heavy ion collisions. 
We have analysed STAR data \cite{Abelev:2007rw} on centrality dependence of $\phi$ mesons multiplicity, mean $p_T$ and (integrated) $v_2$ and obtained an accurate estimate of QGP viscosity over entropy ratio. Details can be found in \cite{Chaudhuri:2009uk}.    
With a lattice based EOS, where the confinement-deconfinement transition is a cross-over at $T_c$=196 MeV, hydrodynamic equations are solved with the code AZHYDRO-KOLKATA \cite{Chaudhuri:2006jd,Chaudhuri:2008sj,Chaudhuri:2005ea}.  
Variation of $\phi$ multiplicity in 0-5\% Au+Au collision, with the central energy density of fluid is shown in the left panel of Fig.\ref{F1}. For $\eta/s$=0-0.16, experimental multiplicity (the shaded region) can be fitted by changing the initial energy density. In table.\ref{table1}   the central energy density and temperature  required to reproduce the experimental
multiplicity data are noted.  More viscous fluid require less initial energy density/temperature as entropy is generated during the evolution. One note that with increasing viscosity mean $p_T$ increases and simultaneous fit to experimental multiplicity and mean $p_T$ is
possible only for $\eta/s \leq$0.12. Fit obtained to the STAR data for centrality dependence of $\dn$, $\pt$ and $v_2$,
from hydrodynamic evolution of fluid with viscosity to entropy ratio $\eta/s$=0-0.16 are shown in the right panel. $\chi^2$ analysis indicate that the
best fit to the data is obtained with  $\eta/s=0.07\pm 0.03$ \cite{Chaudhuri:2009uk}.  The central value is very close to the KSS bound, $\eta/s=1/4\pi$ \cite{Kovtun:2003wp}. In \cite{Chaudhuri:2009uk}, systematic uncertainty in $\eta/s$ due to uncertain initial conditions was also studied. Systematic
uncertainty due to uncertain initial time $\tau_i$=0.2-1.0 fm, freeze-out temperature $T_F$=140-160 MeV, percentage of hard scattering contribution in initial energy density $f$=0-95\%, initial transverse velocity $v_r=tanh(\alpha r)$, $\alpha$=0-0.6, inaccuracy in the hydrodynamical code were evaluated to obtain,   $\eta/s=0.07\pm 0.03 (stat. ) \pm 0.14 (sys.)$. We note here that all the sources of systematic error   e.g. uncertainty in initial stress tensors, relaxation time etc. are not included. Their inclusion will increase systematic uncertainty even more.

\subsection{Scaling property of elliptic flow and QGP viscosity}\label{sec5}

Elliptic flow in non-central collisions is a key observable in   establishing fluid like behavior of the medium produced in Au+Au collisions at RHIC.  
Scaling properties of elliptic flow ($v_2$) has been studied by the STAR and PHENIX collaboration  \cite{:2008ed,Taranenko:2007gb}.  Both the collaborations observed 'universal scaling', $v_2/(n_q \langle v_2 \rangle_{ch})$, elliptic flow scaled by the constituent quark numbers and charged particles $v_2$, of different particle species in different collision centrality, scales with $KE_T/n_q$, the transverse kinetic energy per constituent quark number.   Initial eccentricity scaling or the constituent quark number scaling are not indicated in the data.

\begin{figure}[ht]
\begin{minipage} {14pc}
  {In Fig.\ref{F2}, we have studied universal scaling of elliptic flow for $\pi^-$, $K^+$, proton, $\phi$ and $\Omega$ in 0-60\%  Au+Au collisions \cite{Chaudhuri:2009ud}.  Only a few collision centralities are shown for clarity. Contrary to experiments, in hydrodynamic simulations, irrespective of fluid viscosity, elliptic flow does not follow the   universal scaling. 
The solid line in Fig.\ref{F2} is the approximate scaling function   as obtained in the PHENIX experiment \cite{Taranenko:2007gb} (we have ignored the fluctuations of the scaling function). 
  }
\end{minipage}\hspace{0.5pc}
\begin{minipage} {16pc}
\center
 \resizebox{0.7\textwidth}{!}{%
  \includegraphics{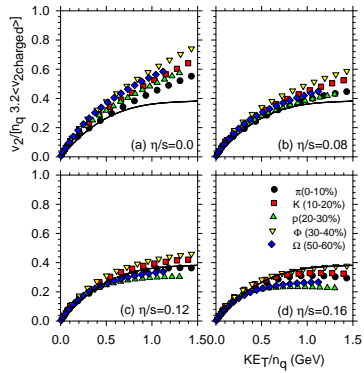}
}
\caption{\label{F2} (color online) Universal scaling in viscous hydrodynamics.}
 \end{minipage} 
\end{figure}

 \noindent 
  
To measure the departure of simulated flows from the experimental 'universal scaling function',  we defined a scaling violation function $F$,

\begin{equation} \label{eq4}
F=\Sigma [(v_{2scaled}^{th}) - v_{2scaled}^{ex}]^2
\end{equation} 

In 0-60\% Au+Au collisions, simulated flows for $\pi^-$, $K^+$, p, $\phi$ and $\Omega$ show minimum departure from universal scaling for  $\eta/s=0.12\pm 0.03$ \cite{Chaudhuri:2009ud}.  

In \cite{Chaudhuri:2009ud} initial eccentricity or constituent quark number scaling was also studied. Initial eccentricity scaling, or constituent quark number scaling is only approximate in ideal and viscous flow, though it appear that scaling property is better satisfied in viscous than ideal fluid flow. A similar trend can be observed in Fig.\ref{F2}.
The result is contradictory to our expectation. Since viscous correction introduces a microscopic scale, it is expected that   any scaling property that would have been observed in ideal fluid, will worsens in  viscous fluid. The issue is discussed in  \cite{Chaudhuri:2009ud}. In viscous evolution, elliptic flow has a contribution from the non-equilibrium part of the distribution function. In hydrodynamic evolution with   lattice based EOS with cross-over transition, the viscous fluxes decreases rapidly. At freeze-out fluid behave more like an ideal fluid. Viscosity however changes the freeze-out surface. Changed freeze-out surface, but small viscous correction, may possibly be the reason for obtaining better scaling in viscous dynamics than in ideal dynamics.

 \begin{table}[ht] 
\caption{\label{table2} Best fitted $\eta/s$ as a function of collision centrality.} 
\begin{tabular}{|c|c|c|c|c|c|c|c|}\hline
coll. centrality&  0-10\% & 10-20\% & 20-30\% & 30-40\% & 40-50\% & 50-60\% & 0-60\%  \\  \hline
$\eta/s$ &$0 \pm 0.03$& $0.051\pm 0.008$ &$0.087\pm0.004$ & $0.109\pm0.003$& $0.134\pm 0.004$ &$0.169 \pm 0.005$ &$0.12 \pm 0.01$ \\ \hline
 \end{tabular} 
 \end{table}  

 \subsection{Centrality dependence of elliptic flow and QGP viscosity}\label{sec6}

\begin{figure}[ht]
\begin{minipage} {14pc}
  {
   Increase of viscosity with collision centrality is not contrary to the present 
  paradigm that $\eta/s$ has a minimum, possibly with a cusp, around the critical temperature  $T=T_c$  \cite{Csernai:2006zz}.   Rather it indicates the increasingly important role of hadronic matter in the development of elliptic flow in peripheral collisions. 
Viscosity to entropy ratio, as obtained from hydrodynamic analysis is averaged over the space-time. Both the QGP and the hadronic phase contribute to the average.  
Schematically, one can write,
   }
\end{minipage}\hspace{0.5pc}
\begin{minipage} {16pc}
\vspace{.2cm}
\center
 \resizebox{0.6\textwidth}{!}{%
  \includegraphics{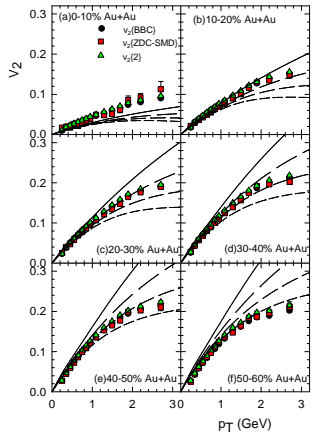}
}
\caption{\label{F3} (color online) Centrality dependence of $v_2$.}
 \end{minipage} 
\end{figure}

PHENIX measurements \cite{Afanasiev:2009wq} for charged particles elliptic flow in 0-60\%  Au+Au collisions are shown in Fig.\ref{F3}. In Fig.\ref{F3}, the solid, dashed, medium dashed and short dashed lines are AZHYDRO-KOLKATA simulations \cite{Chaudhuri:2009hj} for elliptic flow in fluid evolution with (i) $\eta/s$=0 (ideal fluid), (ii) $\eta/s$=0.08, (iii) $\eta/s$=0.12, and (iv) $\eta/s$=0.16, respectively.  In
mid-central or in peripheral collisions, elliptic flow is over predicted in ideal fluid evolution. Data are better explained in viscous fluid evolution. 
It appear that elliptic flow data in more peripheral collisions demand more viscous fluid \cite{Chaudhuri:2009hj}. 
To be more quantative, from a $\chi^2$ analysis, we obtain the best fitted $\eta/s$ in each collision centrality. They are listed in table.\ref{table2}. In 0-60\% collision centrality,
$\eta/s$ varies between 0-0.17.

\begin{equation} \label{eq6}
\frac{\eta}{s} = (1-f_{HAD}) {(\frac{\eta}{s})}^{qgp}(T_{QGP}) + f_{HAD} {(\frac{\eta}{s})}^{had}(T_{HAD})
\end{equation}

\noindent where $f_{HAD}$ is the fraction of the hadronic matter, ${(\frac{\eta}{s})}^{qgp}(T_{QGP})$  is the viscosity of QGP matter at average temperature $T_{QGP}$ and ${(\frac{\eta}{s})}^{had}(T_{HAD})$ is the viscosity
of the hadronic matter at average temperature $T_{HAD}$. Detail analysis \cite{Chaudhuri:2009hj} indicate that 
compared to a central collision, in a peripheral collision, while
$T_{QGP}$ decreases, $T_{HAD}$ remain approximately same. Centrality dependence of extracted $\eta/s$ can be qualitatively understood as due to increased contribution of hadronic phase and decreased contribution of the QGP phase in peripheral collisions than in a central collisions.  

\vspace{0.5cm}
 \begin{figure}[ht]
\begin{minipage} {15pc}
\center
 \resizebox{0.8\textwidth}{!}{%
  \includegraphics{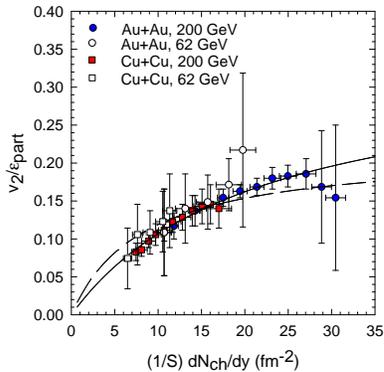}
}
 \end{minipage} \hspace{1pc}
\begin{minipage} {12pc}
\vspace{-1.0cm}
\caption{\label{F4} (color online) PHOBOS data for the centrality dependence of eccentricity scaled elliptic flow in Cu+Cu and Au+Au  collisions at $\sqrt{s}$= 62 and 200 GeV. The solid    line is the fit to the data with Eqs.\ref{eq9} and \ref{eq11}. We have used $\sigma$=3 mb, $K_0$=0.7 and $c_s=\sqrt{1/3}$. The dashed line is the fit in the ideal fluid approximation. Quality of fit is comparatively poor in ideal fluid approximation.}
\end{minipage}
\end{figure}

\section{Knudsen number, hydrodynamic limit for $v_2$ and QGP viscosity}

Ideal hydrodynamics require complete thermalisation. The requirement is relaxed in dissipative hydrodynamics. Incomplete thermalisation in heavy ion collisions can be 
quantified in terms of Knudsen number, $K=\frac{\bar{R}}{\lambda}$, $\bar{R}$ is the characteristic size of the system and $\lambda$ is the mean free path.  
Applicability of   hydrodynamics require that $K^{-1} >>1$. In the Knudsen regime $K^{-1} << 1$ and hydrodynamics is inapplicable. The simple formula,

\begin{equation} \label{eq9}
\left ( \frac{v_2}{\epsilon} \right )^{ex}=\vih
\frac{K^{-1}}{K^{-1}+K_0^{-1}},
\end{equation}

\noindent proposed in \cite{Bhalerao:2005mm} give qualitatively correct behavior of the experimental elliptic flow.  
In the limit of small Knudsen number, experimental flow approaches the ideal hydrodynamic limit $v_2^{ex} \rightarrow v_2^{ih}$, with a correction   linear in $K$. In the other extreme limit of large $K$, $v^{ex}_2 \propto K$. In Eq.\ref{eq9},
$K_0^{-1}$ is a number of the order of unity. In \cite{Bhalerao:2005mm}, Knudsen number was expressed in terms of experimental multiplicity $\frac{dN}{dy}$, 
 
\begin{equation} \label{eq10}
\frac{1}{K}=c_s \sigma \frac{1}{S}\frac{dN}{dy}
\end{equation}

\noindent where   $S$ is a measure of the transverse area of the collision zone,
$c_s$  is the square speed of sound of the medium and $\sigma$ is the inter-particle cross section.  Eq.\ref{eq9} and \ref{eq10} connect two experimental observables, elliptic flow and particle multiplicity  and can be used to determine 
  ideal hydrodynamic limit of elliptic flow $\vih$,  the combination of parameters $K_0\sigma c_s$. They have been used to estimate QGP viscosity from experimental data also. The estimates vary between $\eta/s\approx 0.11-0.27 $
\cite{Drescher:2007cd,Masui:2009pw,Nagle:2009ip}. However, Eq.\ref{eq10} assume isentropic expansion and is invalidated in viscous evolution where entropy is generated. Recently in \cite{Chaudhuri:2010in}, Eq.\ref{eq10} was generalised   to include the effect of entropy generation in viscous evolution, 

\begin{equation} \label{eq11}
\frac{1}{K} \approx  
 \sigma c_s    \left [\frac{1}{S}\frac{dN}{dy} \right ]  
\left [1+\frac{2}{3\tau_iT_i} \left(\frac{\eta}{s}\right)   \right ]^{-3} 
 \end{equation}
 
One immediately observes that neglect of entropy generation during evolution will   over estimate  $K^{-1}$, by the factor $\left [1+\frac{2}{3\tau_iT_i} \left(\frac{\eta}{s}\right)   \right ]^{3}$. 
Eqs.\ref{eq9} and \ref{eq11} can be used to fit experimental data on particle multiplicity and $v_2$. In Fig.\ref{F4}, fit obtained to the PHOBOS data \cite{Back:2004mh,Alver:2006wh} with $\vih=0.33\pm 0.12$, $\etas=0.83\pm 0.51$
is shown. Large uncertainty in $\vih$ and $\etas$ reflect the large uncertainty in PHOBOS data.
Using simulation results of \cite{Chaudhuri:2009uk}, $\etas$ can be converted into more comprehensible viscosity to entropy ratio, $\eta/s=0.17 \pm 0.10 \pm 0.20$.   
   
\begin{figure}[ht]
\begin{minipage}{10pc} 
\resizebox{1.0\textwidth}{!}{\includegraphics{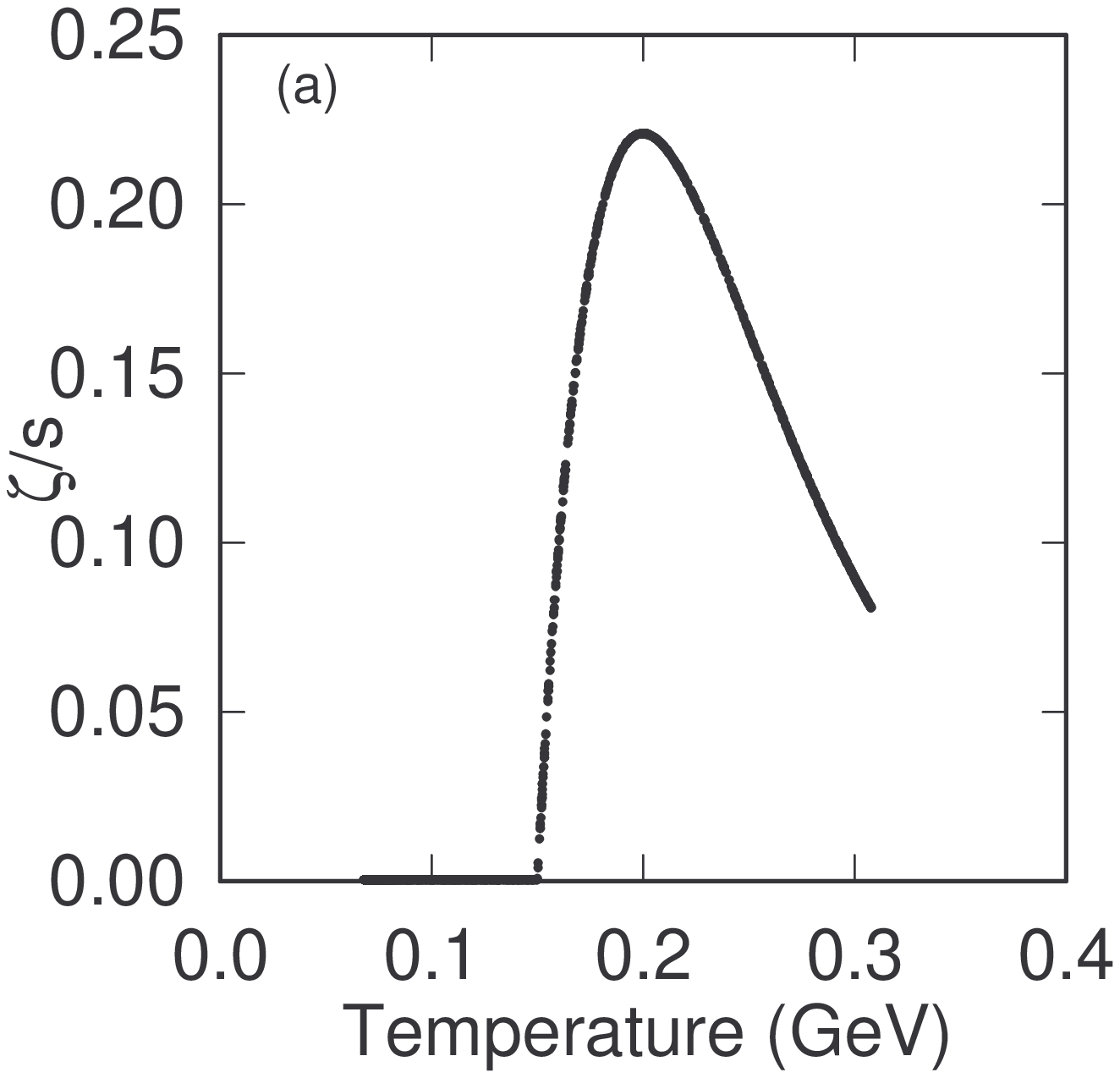}} 
 \end{minipage} 
\begin{minipage}{10pc} 
\resizebox{1.0\textwidth}{!}{\includegraphics{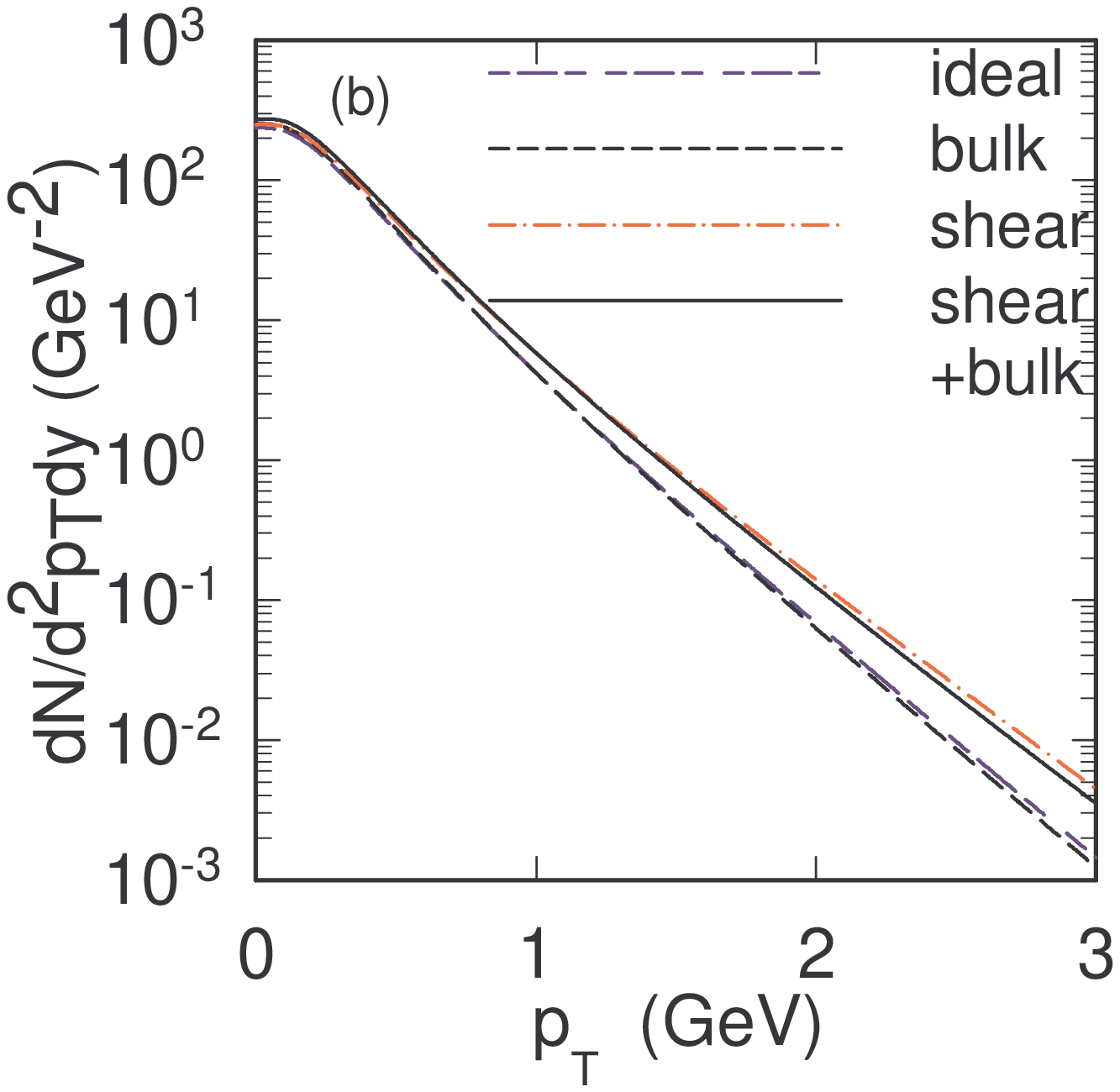}} 
 \end{minipage}
\begin{minipage}{10pc}
\resizebox{0.95\textwidth}{!}{\includegraphics{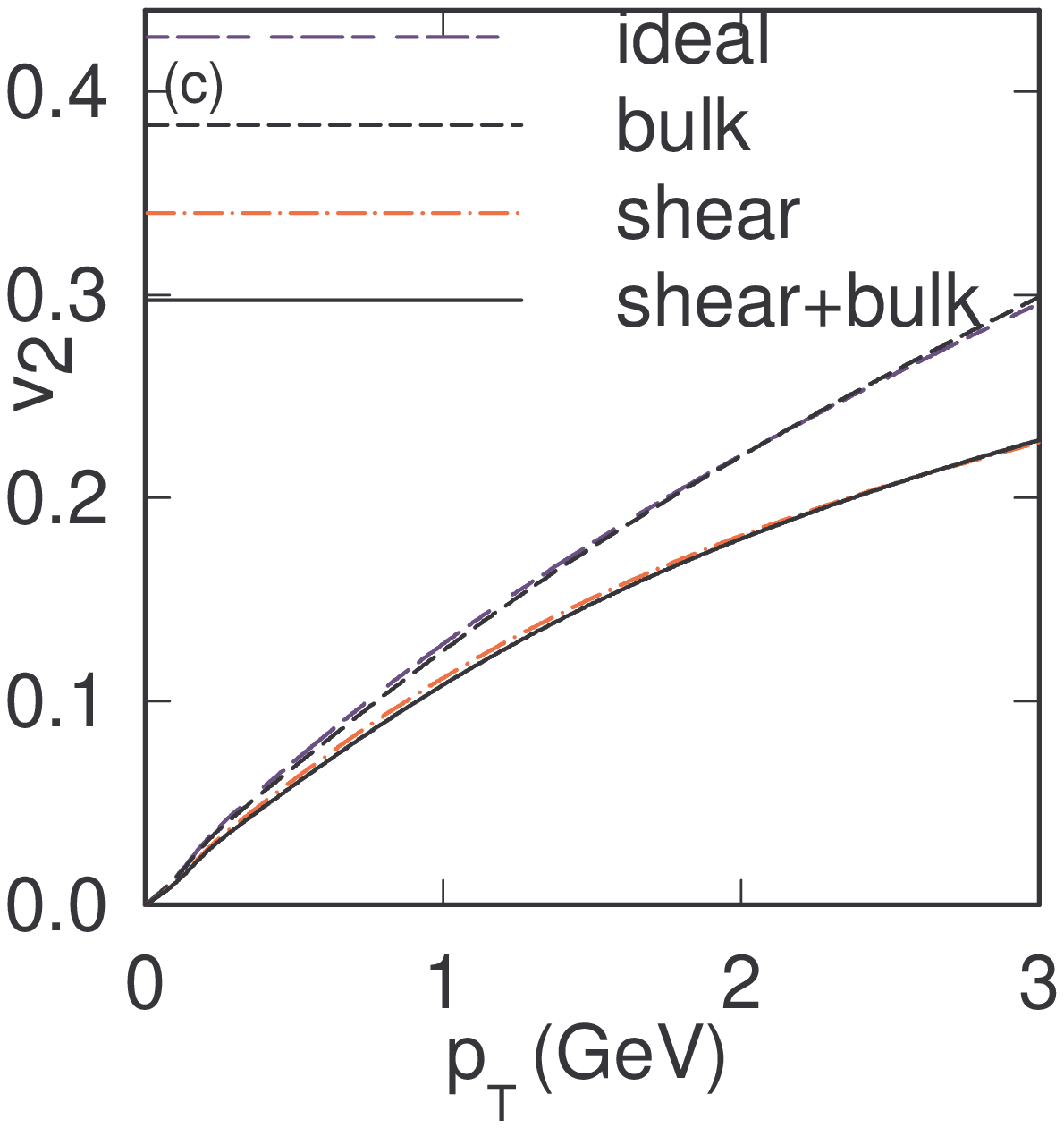}} 
 \end{minipage}
\caption{ \label{F5} (color online) For a parametric bulk viscosity (see left panel) AZHYDRO-KOLKATA simulations for $p_T$ spectra (middle panel) and $v_2$ (right panel) in b=6.5 fm Au+Au collisions. $\eta/s$=0.08.}
\end{figure}
 
 \vspace{-0.5cm}
\section{Bulk viscosity  in heavy ion collisions}

Estimates of $\eta/s$ discussed above ignores the bulk viscosity.
In general bulk viscosity is much smaller than shear viscosity. However, in QCD near the critical point 
bulk viscosity can be   large \cite{Kharzeev:2007wb,Karsch:2007jc}. In \cite{Song:2009rh} effect of bulk viscosity on particle spectra and elliptic flow has been studied in detail. Strong growth of bulk viscosity near the critical point is offset by critical slowing down of the dynamics of bulk pressure, diminishing the bulk pressure contribution to spectra or flow \cite{Song:2009rh}. Recently, we have included the effect of bulk viscosity in AZHYDRO-KOLKATA. Detail results will be published later. In Fig.\ref{F5}, AZHYDRO-KOLKATA simulations for $p_T$ spectra and $v_2$ in b=6.5 fm Au+Au collisions are shown. $p_T$ spectra and $v_2$ are marginally changed when effects of bulk viscosity is included. Indeed, the sensitive observable $v_2$ also show little modification. It is understood. Bulk viscosity do not introduce additional asymmetry in the system and elliptic flow 
is largely unaffected by bulk viscosity.  Estimates of $\eta/s$ obtained by neglecting bulk viscosity will not be altered significantly with inclusion of bulk viscosity.
 
\section{Summary}

To summarise, we have briefly discussed causal dissipative hydrodynamics. Some
recent attempts to extract QGP viscosity from  comparison of hydrodynamical simulations with experimental data are also discussed. Extracted values of viscosity to entropy ratio, from a variety of experimental data, lies within a narrow 
range, $\eta/s \approx$0.08-0.20.  
However, due to uncertain initial conditions in hydrodynamic simulations, systematic error in extracted values of $\eta/s$ could be large,   $\sim$200\% or more. It appear that precise estimate of QGP viscosity from experimental data will not be possible, unless inputs of hydrodynamical models are constrained. 


\end{document}